\documentclass[fleqn,usenatbib]{mnras}

\usepackage{newtxtext,newtxmath}

\usepackage[T1]{fontenc}

\DeclareRobustCommand{\VAN}[3]{#2}
\let\VANthebibliography\thebibliography
\def\thebibliography{\DeclareRobustCommand{\VAN}[3]{##3}\VANthebibliography}

\usepackage{graphicx}	
\usepackage{amsmath}	

\usepackage{scalerel}
\usepackage{tikz}
\usetikzlibrary{svg.path}

\def\chandra    {{\em Chandra}\/}

\def\xmm        {{\em XMM-Newton}\/}

\def\xrism        {{\em XRISM}\/}
\def\athena        {{\em Athena}\/}
\def\nustar        {{\em NuSTAR}\/}
\def\integral        {{\em INTEGRAL}\/}

\definecolor{orcidlogocol}{HTML}{A6CE39}
\tikzset{
  orcidlogo/.pic={
    \fill[orcidlogocol] svg{M256,128c0,70.7-57.3,128-128,128C57.3,256,0,198.7,0,128C0,57.3,57.3,0,128,0C198.7,0,256,57.3,256,128z};
    \fill[white] svg{M86.3,186.2H70.9V79.1h15.4v48.4V186.2z}
                 svg{M108.9,79.1h41.6c39.6,0,57,28.3,57,53.6c0,27.5-21.5,53.6-56.8,53.6h-41.8V79.1z M124.3,172.4h24.5c34.9,0,42.9-26.5,42.9-39.7c0-21.5-13.7-39.7-43.7-39.7h-23.7V172.4z}
                 svg{M88.7,56.8c0,5.5-4.5,10.1-10.1,10.1c-5.6,0-10.1-4.6-10.1-10.1c0-5.6,4.5-10.1,10.1-10.1C84.2,46.7,88.7,51.3,88.7,56.8z};
  }
}

\newcommand\orcidicon[1]{\href{https://orcid.org/#1}{\mbox{\scalerel*{
\begin{tikzpicture}[yscale=-1,transform shape]
\pic{orcidlogo};
\end{tikzpicture}
}{|}}}}

\title[Inverse-Compton emission in a galaxy group]{Discovery of inverse-Compton X-ray emission and estimate of the volume-averaged magnetic field in a galaxy group}

\author[F. Mernier et al.]{
F. Mernier$^{\small\orcidicon{0000-0002-7031-4772}}$,$^{1,2,3,4}$\thanks{E-mail: fmernier@umd.edu}\thanks{\textit{ESA Research Fellow.}}
N. Werner$^{\small\orcidicon{0000-0003-0392-0120}}$,$^{5}$
J. Bagchi$^{\small\orcidicon{0000-0002-2922-2884}}$,$^{6}$
M.-L. Gendron-Marsolais$^{\small\orcidicon{0000-0002-7326-5793}}$,$^{7,8}$
Gopal-Krishna,$^{9}$
\newauthor
M. Guainazzi$^{\small\orcidicon{0000-0002-1094-3147}}$,$^{3}$
A. Richard-Laferri\`{e}re$^{\small\orcidicon{0000-0001-7597-270X}}$,$^{10}$
T. W. Shimwell$^{11}$
and A. Simionescu$^{\small\orcidicon{0000-0002-9714-3862}}$,$^{4,12,13}$ 
\\
$^{1}$NASA Goddard Space Flight Center, Code 662, Greenbelt, MD 20771, USA \\
$^{2}$Department of Astronomy, University of Maryland, College Park, MD 20742-2421, USA\\
$^{3}$European Space Agency (ESA), European Space Research and Technology Centre (ESTEC), Keplerlaan 1, 2201 AZ Noordwijk, The Netherlands \\
$^{4}$SRON Netherlands Institute for Space Research, Niels Bohrweg 4, 2333 CA Leiden, The Netherlands\\
$^{5}$Department of Theoretical Physics and Astrophysics, Faculty of Science, Masaryk University, Kotl\'a\v{r}sk\'a 2, Brno, CZ-611 37, Czech Republic \\
$^{6}$Department of Physics \& Electronics, CHRIST (Deemed to be University), Hosur Road, Bengaluru, 560029, India \\
$^{7}$European Southern Observatory, Alonso de C\'{o}rdova 3107, Vitacura, Casilla, 19001, Santiago de Chile, Chile \\
$^{8}$Instituto de Astrof\`{i}sica de Andaluc\`{i}a (IAA-CSIC), Glorieta de la Astronom\`{i}a, 18008 Granada, Spain \\
$^{9}$UM-DAE Centre for Excellence in Basic Sciences (CEBS), Vidyanagari, Mumbai - 400098, India \\
$^{10}$Institute of Astronomy, Madingley Road, Cambridge CB3 0HA, UK \\
$^{11}$ASTRON, Netherlands Institute for Radio Astronomy, Oude Hoogeveensedijk 4, Dwingeloo, 7991 PD, The Netherlands \\
$^{12}$Leiden Observatory, Leiden University, PO Box 9513, NL-2300 RA Leiden, The Netherlands \\
$^{13}$Kavli Institute for the Physics and Mathematics of the Universe (WPI), University of Tokyo, Kashiwa 277-8583, Japan
}

\date{Accepted 2023 July 7. Received 2023 July 5; in original form 2022 July 19}

\pubyear{2023}

\begin{document}
\label{firstpage}
\pagerange{\pageref{firstpage}--\pageref{lastpage}}
\maketitle

\begin{abstract}
Observed in a significant fraction of clusters and groups of galaxies, diffuse radio synchrotron emission reveals the presence of relativistic electrons and magnetic fields permeating large scale systems of galaxies. Although these non-thermal electrons are expected to upscatter cosmic microwave background photons up to hard X-ray energies, such inverse-Compton (IC) X-ray emission has so far not been unambiguously detected on cluster/group scales. Using deep, new proprietary \xmm\ observations ($\sim$200~ks of clean exposure), we report a 4.6\,$\sigma$ detection of extended IC X-ray emission in MRC 0116\,+111, an extraordinary group of galaxies at $z = 0.131$. Assuming a spectral slope derived from low frequency radio data, the detection remains robust to systematic uncertainties. Together with low frequency radio data from the Giant Metrewave Radio Telescope (GMRT), this detection provides an estimate for the volume-averaged magnetic field of $(1.9 \pm 0.3)$~$\mu$G within the central part of the group. This value can serve as an anchor for studies of magnetic fields in the largest gravitationally bound systems in the Universe. \\
\end{abstract}





\section{Introduction}
\label{sec:intro}


At the dawn of the high energy astrophysics era, the origin of the bright, extended X-ray emission seen towards clusters and groups of galaxies was debated \citep{brecher1972,lea1973}. The advent of dedicated space observatories established that these systems shine in X-rays via thermal bremsstrahlung and line emission from a hot ($10^{7-8}$~K), collisionally ionized medium permeating them \citep{sarazin1986,bohringer2010}. However, radio wavelength detections of diffuse synchrotron radiation from relativistic electrons in some clusters \citep[e.g.][]{vanweeren2019} implies the existence of a widespread relativistic plasma which should also upscatter cosmic microwave background (CMB) photons up to X-ray energies via inverse-Compton (IC) scattering. 

The importance of detecting this diffuse IC X-ray emission in clusters/groups of galaxies has been long highlighted \citep{petrosian2008,feretti2012} because in combination with radio observations, it can yield the most direct estimate of the intracluster magnetic field. The latter is, in fact, a critical input for understanding the origin and evolution of cluster magnetic fields, which is fragmentary at the moment \citep{vazza2021}. Other methods for estimating magnetic fields in clusters and groups of galaxies are afflicted by observational biases. The estimates based on the well-known technique of Faraday rotation \citep[e.g.][]{govoni2004,bohringer2016} depend critically on the local topography of magnetic fields along the line of sight and, furthermore, remain highly sensitive to the foreground Galactic interstellar dust. The alternative method utilizes the radio synchrotron emission, which depends on the intensity of magnetic fields and the energy density of relativistic electrons. Usually, these two parameters are estimated by assuming equipartition between magnetic and particle energy densities---whereby the total energy density is minimized \citep[e.g.][]{feretti2012}. The validity of this assumption on various physical scales, however, remains beset with formidable uncertainty \citep[e.g.][]{pfrommer2004,petrosian2008}. The degeneracy inherent to the equipartition assumption can be effectively broken using the detection of IC X-ray emission, enabling a reliable estimate of the energy density of the relativistic electron population.

Measuring diffuse IC emission at large scales thus constitutes a promising way to boost our understanding of cosmic magnetic fields. From the earliest hunts \citep{rephaeli1987,rephaeli1988} to recent attempts over the past three decades \citep{wik2012,ota2014,cova2019,rojas2021} observations have only yielded upper limits for any diffuse IC X-rays from clusters. The reported detections pertain to the Coma cluster \citep{rephaeli1999,rephaeli2002,fusco1999,fusco2004}, Abell\,85 \citep{bagchi1998}, the Ophiuchus cluster \citep{eckert2008}, and the Bullet cluster \citep{petrosian2006,ajello2010}; however all these claims have been found spurious (or rendered controversial) in subsequent more sensitive observations \citep{rossetti2004,fusco2007,eckert2007,lutovinov2008,durret2005,wik2009,wik2014}. A key drawback with these objects is that the intracluster gas at such high temperatures radiates thermal X-rays that dominate the IC component even at a few keV energies, rendering the latter practically undetectable. This circumstance often mandates complementary observations at higher energies (i.e. even beyond 10~keV), for example using the \nustar\ or \integral\ space observatories, in order to confirm the putative IC X-rays. However, this often gets mired in imperfect cross-calibration between the instruments covering different energy bands, which adds substantial uncertainty. Moreover, the hot gas pervading these dynamically disturbed clusters often has a rich multitemperature structure, particularly due to recently shocked regions \citep{donnert2017}, which complicates spectral modelling.

An excellent candidate having the potential to circumvent these difficulties is the galaxy group MRC\,0116+111 \citep[or OTL\,0116+111 at its discovery;][]{joshi1980}. In addition to its relatively cool gaseous medium \citep[$kT \simeq 0.7$--0.8~keV;][]{mernier2019}, this group is known to be a source of spectacularly bright diffuse radio emission \citep{gopal2002,bagchi2009}, consisting of a pair of $\sim$50\,kpc wide features of enhanced brightness surrounded by a more extended diffuse emission spanning nearly 2\,arcmin. Rather than confined, well structured lobes, the broad shape of the inner, double-peaked radio structure at both moderate and higher spatial resolution \citep{bagchi2009} suggests lobe \textit{remnants}, likely arguing for a rather \textit{ancient} nuclear activity in the galaxy situated between them. This is supported by the fact that no active radio nor X-ray point source is detected within the system \citep{mernier2019}. The origin of the entire extended structure, however, is still unclear. First thought to be a mini-halo \citep[i.e. resulting from a turbulent cool core induced by sloshing; ][]{bagchi2009}, the large radio extent compared to its spatially confined X-ray counterpart suggests an alternative origin for its radio emission \citep{mernier2019}. Its interpretation as `giant' radio halo (i.e. resulting from a merger between two galaxy systems) also seem unlikely as the latter have not been unambiguously detected at group scales so far---probably because the level of turbulence needed to re-accelerate relativistic fossil electrons is much weaker in such low-mass systems \citep[for a few radio halo detections in intermediate mass systems, however, see][]{botteon2021,paul2021}. A more likely scenario makes this radio source possibly related to an ancient pair of relativistic plasma bubbles (hence, linking both the inner lobe remnants with the more diffuse surrounding structure), and potentially originating from previously intense activity in the supermassive black hole of its central brightest galaxy \citep[for an extensive discussion, see][]{mernier2019}. How exactly such an episode could have (re-) accelerated electrons over the extended and diffuse structure seen in several radio bands remains an open question \citep{bagchi2009}. Upcoming radio observations with the Low Frequency Array (LOFAR) and JVLA will help to better understand the true nature of this radio source (Richard-Laferri\`{e}re et al., in preparation).

The combination of these two extreme properties---rather low gas temperature and bright diffuse radio emission---makes this group a prime target for detecting diffuse IC X-ray emission. Specifically, the emission spectrum of its thermal plasma should peak at soft X-ray energies, with negligible emission above a few keV, where the non-thermal IC emission, shaped spectrally as a power law, is hence expected to dominate. Motivated by this, and by the fact that magnetic fields are even less well-known in groups than in clusters \citep[see however][]{nikiel2017,nikiel2020}, we report in this paper new results from a deep, dedicated  \xmm\ observation of MRC\,0116+111. This work follows a pilot study using a fairly limited exposure of the same system \citep{mernier2019}.

Section~\ref{sec:data_reduction} presents our data reduction as well as the results from spectral and spatial analysis. Section~\ref{sec:systematics} focuses on the robustness of our results while Sect.~\ref{sec:discussion} discusses their interpretation. Section~\ref{sec:conclusion}~concludes this work. Throughout this study, we assume that $\rm H_{0} = 70$ km s$^{-1}$ Mpc$^{-1}$, $\Omega_{\rm M} = 0.27$ and $\Omega_{\rm \Lambda} = 0.73$.
At the distance of MRC\,0116+111 ($z = 0.131$), 1~arcmin corresponds to 179.0~kpc. Chemical abundances are given in the (proto-) solar units of \citet{lodders2009}. Unless mentioned otherwise, the quoted uncertainties are 1\,$\sigma$.


\section{Data and results}\label{sec:data_reduction}


\subsection{\xmm\ data reduction}

This work is based on the three \xmm\ European Photon Imaging Camera (EPIC) observations of MRC\,0116+111 available so far---taken respectively in 2014 January (ObsID: 0722900101) and December 2020 (ObsID: 0864110101, 0864110201)---for a total raw exposure of $\sim$290~ks. The data reduction process follows the pilot study of \citet{mernier2019} and uses the \xmm\ SAS (Science Analysis System) software (v18.0.0) with the up-to-date calibration files (2021 December). Following the standard pipeline recommendations, the EPIC MOS (1 and 2) and EPIC pn data are processed using, respectively, the commands \texttt{emproc} and \texttt{epproc}. For each of the nine datasets, we carefully identify flaring events and discard them following the 2$\sigma$-clipping method described in \citet{mernier2015}, leaving us with 219.9 and 170.5~ks of clean MOS and pn exposures, respectively.

\subsection{Spectral analysis}\label{subsec:spectral}

Aiming to analyze spectrally over the entire extent of the diffuse radio emission (to maximise the statistics), we extract data from an elliptical region similar to that of \citet[][see also Appendix~\ref{app:regions}]{mernier2019}. The redistribution matrix file (RMF) and ancillary response file (ARF) are obtained from the same elliptical region. Due to the fairly small X-ray extent of MRC\,0116+111 ($\lesssim 1.2$~arcmin), the background can be, and is recommended to be, directly subtracted from our spectra (Sect.~\ref{subsec:background}). We choose a background region (i) located on the same chip as the source region for all instruments used in our observations, (ii) devoid of discrete X-ray sources, and (iii) of the same size and shape as the source region (Fig.~\ref{fig:regions}). The small relative separation between this background region and the source ($\sim 1.6$~arcmin) ensures vignetting effects to be negligible (i.e. less than 5\%). The outcome of this method has been already discussed in \citet{mernier2019}, however it is further addressed below.

We use the SPEX fitting package to fit most of our data in the 0.55--5~keV energy band (to minimize further contamination from the soft foreground and the hard instrumental background) with the C-statistics fitting method \citep{kaastra2017}. Spectral bins are grouped optimally \citep{kaastra2016}, though with a gradually wider rebinning with increasing energy (within factors 5--80) to ensure that the count rate of every bin remains significantly above zero. The latter is also necessary to avoid possible biases from C-statistics in low counts, background subtracted fits (see the SPEX manual).

Our modelling consists of a collisionally ionized plasma emission model  (i.e. \texttt{cie}, thermal model) and a power law model (i.e. \texttt{po}; non-thermal emission model), both redshifted \citep[$z = 0.131$;][]{mernier2019} and absorbed by atomic interstellar hydrogen \citep[$n_\mathrm{H} = 3.81 \times 10^{20}$~atoms~cm$^{-2}$;][]{kalberla2005}. Following its close to universal value, the metallicity ($Z$) in the thermal model is fixed to 0.3~Solar with all relevant abundances tied to the Fe value \citep{urban2017,mernier2018c}. Under the assumption of an IC origin for the non-thermal component, the slope $\Gamma$ of the \texttt{po} model is determined by the radio spectral index $\alpha_\mathrm{syn}$ of the source below its spectral break observed near $\sim$400~MHz \citep[$\alpha_\mathrm{syn} = 0.55 \pm 0.05$;][]{bagchi2009}, and is thus fixed to $\Gamma = 1.55$ \citep[see also][]{mernier2019}. This assumption is further tested and discussed in Sect.~\ref{subsec:gamfr} (see also Appendix~\ref{app:gamfr}).
Found with a best-fitting temperature of $kT = (0.64 \pm 0.03)$~keV, the pure thermal model systematically leaves an X-ray excess in the 2--5~keV band, which is well-reproduced by a non-thermal component which is detected at a high $7.1\,\sigma$ significance level. 

\begin{figure}
\centering
	\includegraphics[angle=0,width=0.47\textwidth, trim={0cm 0cm 0cm 0.8cm},clip]{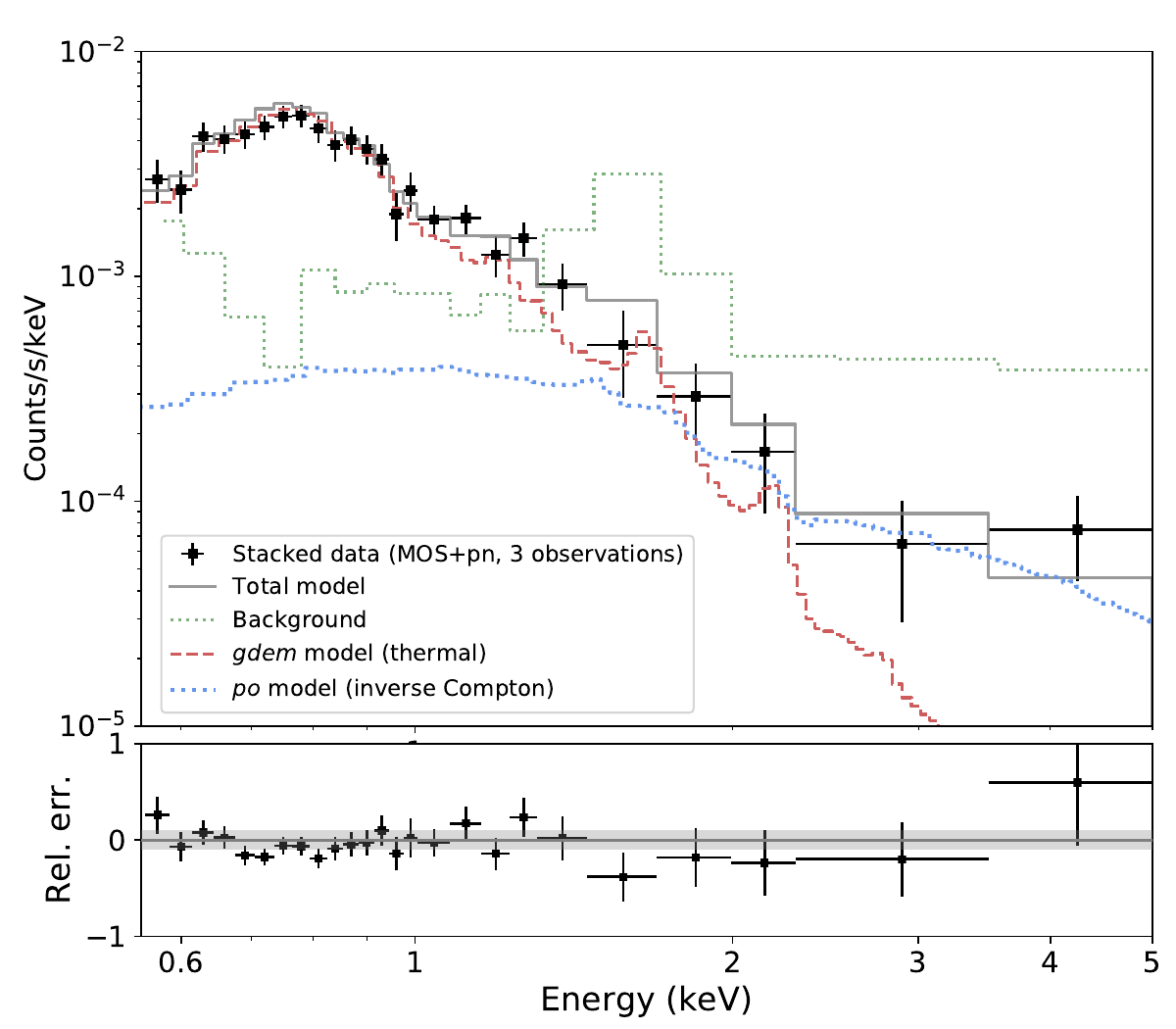}
\caption{Combined, background subtracted EPIC MOS and pn spectrum of MRC\,0116+111 (stacked and spectrally rebinned here for clarity).   The additive thermal (\texttt{gdem}) and non-thermal components are displayed in unbinned resolution for visibility. Spectral residuals to the total model (i.e. [data - model] / model) are shown in the bottom panel.
}
\label{fig:spectrum}
\end{figure}

Multitemperature structure of the gas expectedly extends the spectral profile of the thermal component to higher energies as compared to the oversimplistic assumption of a single-temperature intragroup plasma. As a more realistic modelling, we reiterate the spectral analysis replacing our \texttt{cie} component with a Gaussian-shaped multitemperature distribution of width $\sigma_T$ fixed to 0.2 \citep[i.e. \texttt{gdem}, following the average distribution for other groups;][]{mernier2016a}. Figure~\ref{fig:spectrum} shows our background subtracted integrated spectrum along with this more realistic, multitemperature modelling (\texttt{gdem+po}). We find a similar average best-fitting temperature of $kT = (0.62 \pm 0.04)$~keV as for the single-temperature modelling, with the non-thermal component (blue) systematically dominating over the thermal component (red) beyond $\sim$2~keV and still being detected at $4.6\,\sigma$. Additional multitemperature models are discussed in Sect.~\ref{subsec:multiT}. Although the background dominates in the hard band, its shape is flat and the emission from the group contributes to at least 15--20\% beyond 2~keV, still allowing to place strong constraints on the hard tail, given our excellent statistics. Uncertainties related to the background are discussed further in Sect.~\ref{subsec:background}.

\subsection{Spatial analysis}\label{subsec:spatial}

Motivated by the presence of a possible non-thermal component dominating the X-ray emission beyond $\sim$2~keV, we now analyze its spatial distribution. With the notable exception of two bright sources $\sim$80~arcsec NE and SW away from the group (thus outside of our region of interest), the search for point-like sources (performed via the routine \texttt{edetect\_chain}) leads to no significant detection in the vicinity of MRC\,0116+111. The brighter of the two point sources is in fact a known background quasar \citep[SDSS\,J011904.92\,+112420.4; $z=1.0$;][]{lyke2020} clearly unrelated to the group.

\begin{figure*}
\centering
	\includegraphics[angle=0,width=0.77\textwidth, trim={0cm 0.1cm 0cm 0.25cm},clip]{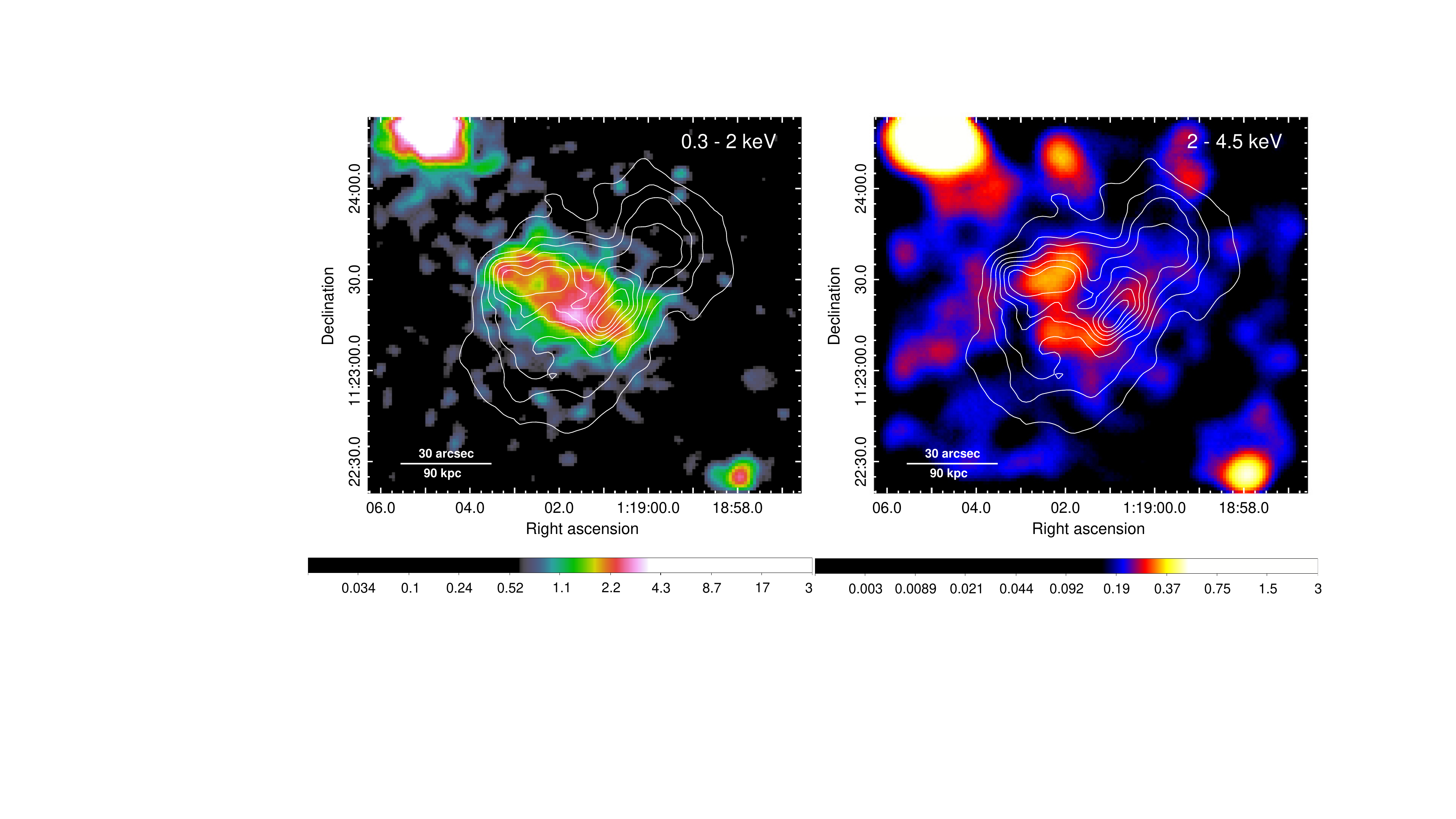}
\caption{Stacked, background subtracted, vignetting-corrected \xmm\ EPIC surface brightness images of MRC\,0166+111, overlayed with radio contours (621~MHz, GMRT).
\textit{Left:} Soft band (0.3--2~keV), smoothed with a Gaussian of radius 4.
\textit{Right:} Hard band (2--4.5~keV), smoothed with a Gaussian of radius 9. Units of the color bars are net counts corrected from vignetting (hence, slightly larger than the true number of counts in the off-axis directions). The negative counts resulting from background subtraction have been artificially set to zero.
}
\label{fig:EPIC}
\end{figure*}

\begin{figure*}
\centering
	\includegraphics[angle=0,width=0.70\textwidth, trim={0cm 0cm 0cm 0cm},clip]{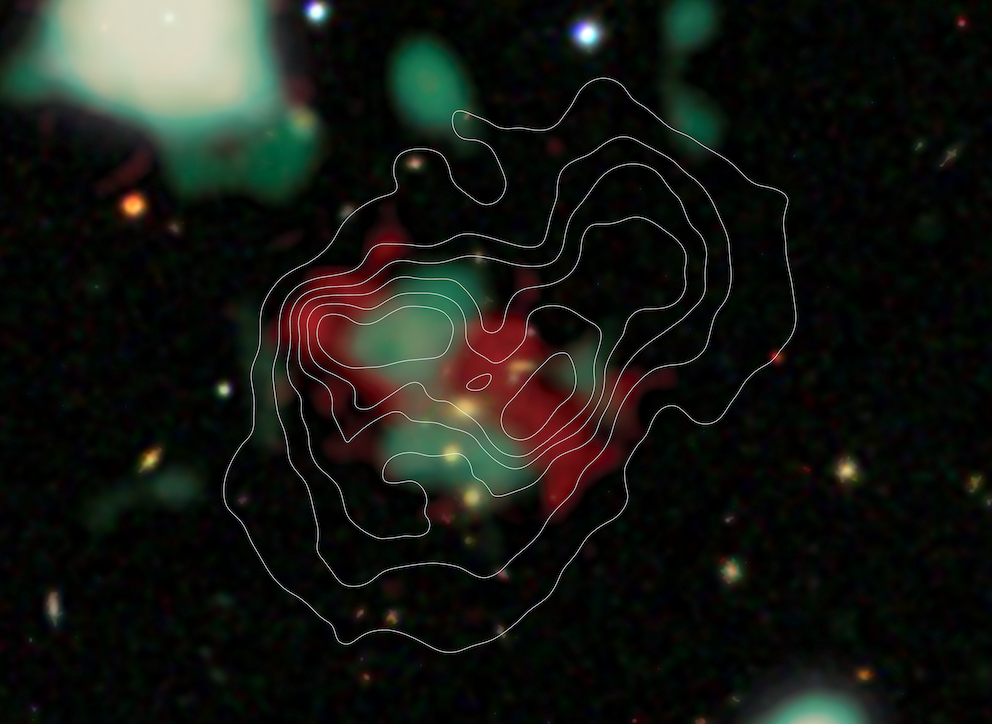}
\caption{Composite, multiwavelength image of MRC\,0116+111. The group's galaxies, seen in optical (SDSS in z, i, r, g, and u bands) are overlaid with the soft (red) and hard (turquoise) X-ray emission, interpreted as the emission of the intra-group thermal gas and IC emission of relativistic electrons. The white contours show the diffuse radio emission at 621~MHz using GMRT. 
}
\label{fig:composite}
\end{figure*}

Figure~\ref{fig:EPIC} shows that, unlike its soft thermal counterpart (left), the diffuse hard X-ray emission (right) has a slightly elongated peak that coincides strikingly with the eastern peak of radio synchrotron emission imaged at 621~MHz \citep{bagchi2009}. We also note the presence of a second X-ray peak of similar projected size and morphology $\sim$60~kpc south of the above-mentioned peak. Beyond this hard X-ray double-peaked region, a more diffuse component can be seen extending across $\sim$130~kpc, as further discussed below. Figure~\ref{fig:composite} shows a multiwavelength composite image of the group, further highlighting the spatial differences between the soft and the hard X-ray components as well as a comparison with the member galaxies and the diffuse radio emission. 

\begin{figure*}
\centering
	\includegraphics[angle=0,width=0.33\textwidth, trim={0cm 0cm 0cm 0cm},clip]{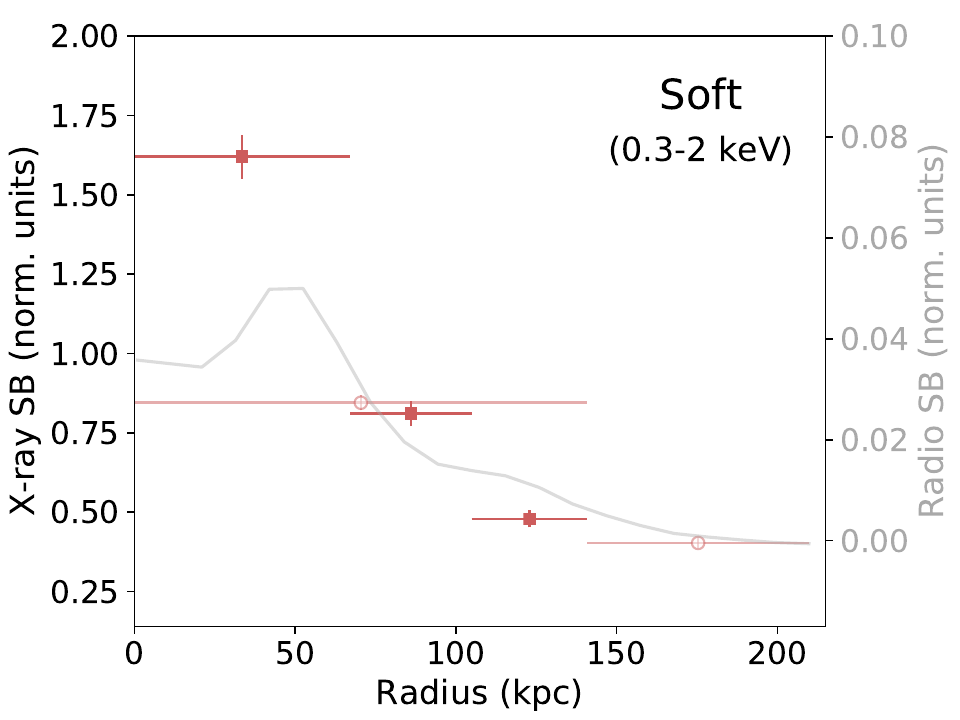}
	\includegraphics[angle=0,width=0.33\textwidth, trim={0cm 0cm 0cm 0cm},clip]{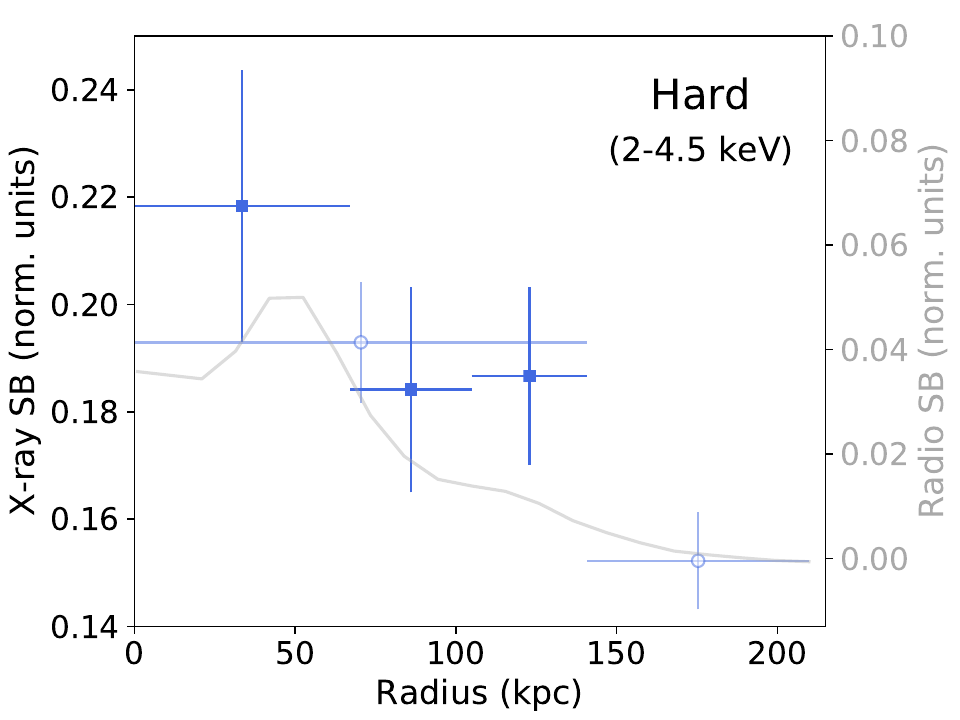}
	\includegraphics[angle=0,width=0.33\textwidth, trim={0cm 0cm 0cm 0cm},clip]{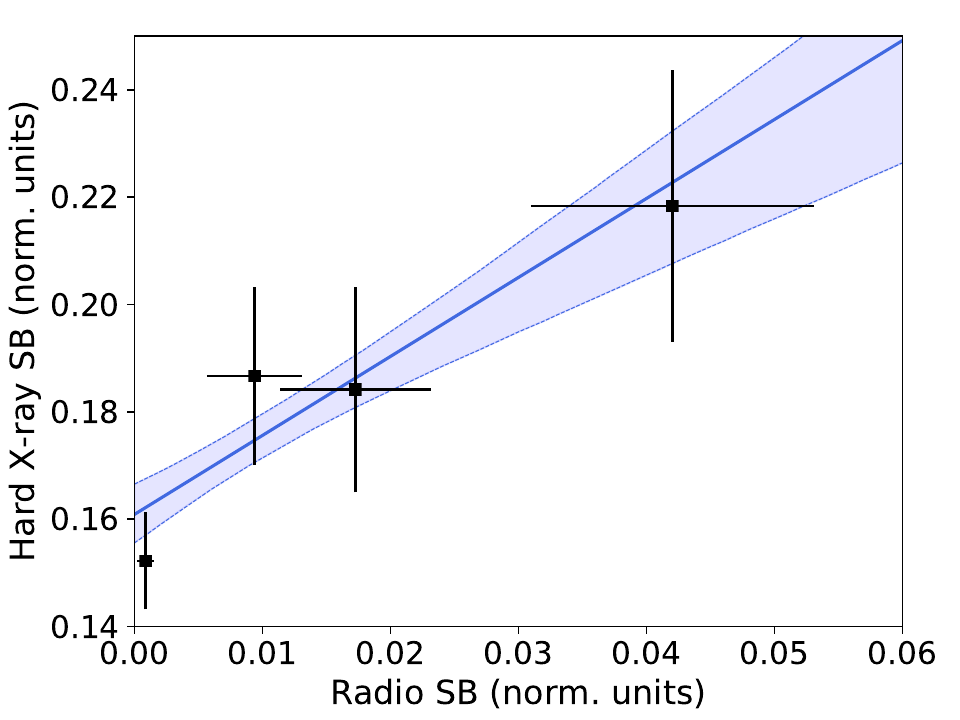}
\caption{\textit{Left and middle panels:} Surface brightness (SB) profile of the soft (0.3--2~keV; \textit{left}) and hard (2--4.5~keV; \textit{middle}) X-ray emission of MRC0116\,+111 along the wedge region defined in Fig.~\ref{fig:regions} (hence excluding the double-peaked region seen in hard X-rays). As explained in the text, we use successively a finer and a coarser binning (i.e. two and four radial bins along the wedge region, respectively). The outermost section of the wedge is largely dominated by the background. For comparison, the radio SB profile along the same region is also displayed. \textit{Right panel:} Correlation between radio and hard X-ray brightness for the regions investigated in this SB analysis. The straight line shows the best-fit linear regression model and the envelope indicates its 32\% and 68\% percentiles.
}
\label{fig:SB}
\end{figure*}

 To further verify the peculiar spatial configuration of these two X-ray components, we extract from the soft and hard X-ray images (Fig.~\ref{fig:EPIC}) surface brightness (SB) profiles along a wedge region centred on the X-ray core and extending beyond the detected radio emission at 621~MHz (Fig.~\ref{fig:regions}). This region is defined so as to exclude the double-peaked region seen in the hard image, although it remains well-suited for the soft band as well (which shows a more elliptical extent; Fig.~\ref{fig:EPIC} left). The outer wedge, falling outside of any apparent X-ray and radio emission, represents the surrounding background emission (with similar SB to the larger background elliptical region to within $<7\%$). Investigated over four radial bins (Fig.~\ref{fig:SB} left and middle panels), we find a signal in both soft and hard bands that clearly extends out to at least $\sim$130~kpc away from the core. In fact, when performing our analysis over two regions, inside and outside the radio extent, the hard X-ray diffuse region is brighter than the outer background region at almost $3\,\sigma$. Although the coarse spatial binning somewhat limits our interpretation, it is interesting to note that the profile of this hard X-ray diffuse component apparently correlates with the radio diffuse emission. This is shown further in the right panel of Fig.~\ref{fig:SB}, where our four radial bins seem to scale linearly in their radio and hard X-ray SB, with a correlation coefficient measured to be $\rho = 0.8 \pm 0.3$ (taking all error bars into account). We have also verified that all these results remain virtually unchanged when the apparent hard X-ray ``clump'' north of the western radio peak (seen in pale red on Fig.~\ref{fig:EPIC} right) is masked out from our analysis. Although the limited X-ray photon statistics does not allow us to replicate this analysis for other quadrants surrounding the double-peaked region, this analysis clearly demonstrates the existence of a truly diffuse hard X-ray component extending well outside the central galaxy and matching the synchrotron radio emission.

\subsection{Resolved and unresolved point sources}\label{subsec:point_sources}

Except one (early-type) galaxy apparently within the southern hard X-ray peak, no other optical counterpart coincides with the extent of the hard X-ray emission. Across the entire field of view, the weakest detected point source has a 2--8~keV flux of $2.2 \times 10^{-15}$~erg~cm$^{-2}$~s$^{-1}$. Extrapolating this value from the $\log\,N$--$\log\,S$ distribution of unresolved point sources from Chandra deep surveys \citep{hickox2006,lehmer2012}, the integrated emission measure contributed by unresolved discrete sources can be estimated for our observations and is found to be $3.7 \times 10^{49}$~photons~s$^{-1}$~keV$^{-1}$. This value is further used in Sect.~\ref{subsec:background} to model the cosmic X-ray background (CXB). Rescaling this value on the hard X-ray double-peaked region, we find that the hard tail exceeds the estimated CXB by a factor $\sim$5, hence making unresolved point sources an unlikely explanation. In addition, a short ($\sim$18~ks) snapshot observation taken by us with \chandra\ ACIS confirms the lack of any major contribution from point sources within the entire region encompassed by the radio emission. In fact, the only two point source candidates tentatively detected in the ACIS broad-band image (0.3--7~keV) account for less than $\sim$6\% and $\sim$11\% of the total flux from the region \citep{mernier2019}. More specifically, the faintest ACIS point source is detected with a 2--8~keV flux of $8.62 \times 10^{-17}$~erg~cm$^{-2}$~s$^{-1}$. Recovering the hard emission of the two central peaks under this scenario would then require at least 33 \chandra-undetected point sources to be concentrated in such a small region which, again, seems highly improbable.

Finally, we have verified that each of the two hard X-ray peaks (each $\sim$25~arcsec across) clearly exceeds the point spread function (PSF) of the EPIC instruments. The apparent extension of the background quasar is naturally explained by its saturated brightness contrast on the figures chosen for better visibility of the group's X-ray emission---even though the quasar's emission profile is found to be perfectly consistent with the shape of the EPIC PSF (and to drop to a negligible level well before the group's X-ray emission). We further note that this quasar is the only point source detected confidently from the ACIS snapshot in the vicinity of the group ($\sim$5\,$\sigma$ in the hard band). Altogether, this makes us confident that the hard X-ray emission of the group is, indeed, significantly extended.

\section{Possible systematics affecting the detection}\label{sec:systematics}


In the previous Section, we have reported the existence of a hard ($\gtrsim 2$~keV) tail in the integrated \xmm\ EPIC spectra of the galaxy group MRC\,0116+111. Remarkably, the spatial distribution of the inferred hard component is also seen to decouple from that of the soft X-ray component (the latter being associated with the thermal gas permeating this poor group).

Given the spectacularly bright, diffuse synchrotron emission observed at radio wavelengths from this group, diffuse IC emission is expected to emerge in the X-ray band and, at first glance, would naturally explain our detected hard X-ray excess.
In order to verify this scenario, however, it is essential to (i) ensure that any systematic effects are smaller than the statistical uncertainties and (ii) explore alternative astrophysical explanations. Both these points can be addressed simultaneously by performing a series of alternative fitting tests, in which the flux density and significance of the IC emission can be directly compared to a simple \texttt{cie+po} modelling case. Whilst the modelling ultimately adopted in this work is that of a multitemperature situation, choosing a single-temperature approach as formal baseline modelling is indeed preferred here to better isolate systematic effects and understand them with minimal interdependence. Table~\ref{tab:systematics} summarizes our tests and their outcomes (see also additional discussion in Sect.~\ref{subsec:spatial_discussion}).

\begin{table}
\begin{centering}
\caption{Investigation of various systematic effects and their impact on the IC detection confidence, defined here as the significance of the flux of the \texttt{po} component in our fits (see text).} \label{tab:systematics}
\label{table:systematics}
\begin{tabular}{lccr}
\hline
\hline
{Description} & {C-stat/d.o.f.} & {$F_\mathrm{IC, 2~keV}$} & {IC det.} \\ 
 &  & {($10^{-15}$~erg/s/cm$^{2}$/keV)} & {confid.} \\ 
\hline  
Single-T: \texttt{cie}				&	278.1/200		&	$1.55 \pm 0.22$	&	$7.1\,\sigma$\\
\hline
Multi-T: \texttt{gdem} 	&	262.5/200		&	$1.30 \pm 0.28$	&	$4.6\,\sigma$ \\
Multi-T: \texttt{wdem} 	&	273.9/200		&	$1.50 \pm 0.22$	&	$6.8\,\sigma$ \\
Multi-T: \texttt{cie+cie} 	&	255.9/198		&	$1.18 \pm 0.27$	&	$4.3\,\sigma$ \\
MOS only 			&	180.1/139		&	$1.27 \pm 0.27$	&	$4.7\,\sigma$ \\
pn only 				&	94.7/52		&	$1.85 \pm 0.34$	&	$5.4\,\sigma$ \\
Modelled bkg 			&	173.9/193		&	$1.76 \pm 0.49$	&	$3.6\,\sigma$ \\
Fit: [0.55--4]~keV		&	274.7/200		&	$1.43 \pm 0.22$	&	$6.4\,\sigma$ \\
Fit: [0.55--10]~keV		&	332.6/209		&	$1.37 \pm 0.21$	&	$6.4\,\sigma$ \\
SPEXACT v2.05		&	278.7/200		&	$1.63 \pm 0.23$	&	$7.2\,\sigma$ \\
AtomDB v3.0.9		&	459.2/404		&	$1.33 \pm 0.22$	&	$6.0\,\sigma$ \\
Free $Z$ and $n_\mathrm{H}$		&	260.5/198	&	$0.92 \pm 0.26$	&	$3.5\,\sigma$  \\
Free $\mathrm{\Gamma}$	&	263.7/199	&	$1.31 \pm 0.22$	&	$6.1\,\sigma$  \\
Free $\Gamma$,  [0.8--5]~keV	&	175.9/129	&	$1.11 \pm 0.53$	&	$2.1\,\sigma$  \\
`Central' region only		&	261.5/197		&	$0.64 \pm 0.09$	&	$7.0\,\sigma$ \\
`Diffuse' region only		&	201.8/138		&	$1.09 \pm 0.22$	&	$5.0\,\sigma$ \\

\hline
\end{tabular}
\par\end{centering}
\end{table}

\subsection{Temperature structure of the thermal emission}\label{subsec:multiT}

First, in principle a hard tail seen in cluster/group spectra could be (at least partly) explained by an oversimplified thermal modelling of its gas component \citep{cova2019}. Astrophysically speaking, the possibility of the gas being heated by shocks---arising from either group mergers \citep{osullivan2019} or active galactic nucleus (AGN) feedback \citep{jetha2008,randall2011,randall2015}---motivates this consideration. To test this hypothesis, in addition to the \texttt{gdem} fit discussed above, we replace the \texttt{cie} model from our initial fit by, successively (i) a power law-distributed multitemperature model (\texttt{wdem}) with its slope $p_T$ fixed to 0.25 \citep[as found, e.g., in the Virgo cluster;][]{kaastra2004}; and (ii) a two-temperature model (\texttt{cie+cie}) with temperatures and emission measures as free variables. Table~\ref{tab:systematics} shows that, in both cases, the non-thermal X-ray emission remains significant at more than $4\,\sigma$. The \texttt{cie+cie} modelling leads to a cooler component with $kT_\mathrm{1} = (0.42 \pm 0.05)$~keV and a hotter component, with an emission measure about half that of the cooler component over our fitting range, with $kT_\mathrm{2} = (1.00 \pm 0.09)$~keV. Importantly, neither of these two modelling attempts can reproduce the hard excess seen beyond 2~keV. We note that adding a third \texttt{cie} component to our \texttt{cie+cie} modelling better reproduces the very soft band only ($kT_\mathrm{3}$ collapsing to the lowest 0.3~keV limit allowed in the fit), resulting in similar fitting statistics and leaving the hard X-ray excess unaltered.  In fact, we note that the only way to reproduce the latter is by using a \texttt{gdem} model with an exceedingly broad temperature distribution width ($\sigma_T \simeq 0.6$). Not only has such an extreme width never been reported in the literature, it would also imply that half of the group's emission originates from a gas component with $kT > 5$~keV, extending even further to $kT = 10$~keV and beyond. Obtaining such a broad range of temperatures from violent mergers and/or AGN feedback would require exceptionally strong shocks with Mach number $\mathcal{M} \gtrsim 5$, which have never been observed in the intracluster medium so far \citep[$\mathcal{M} < 2-3$; e.g.][]{jetha2008,randall2015,vanweeren2019}. Such a broad multitemperature distribution thus seems hardly physical for a group having so few galaxies and with such a modest X-ray luminosity \citep{bagchi2009,mernier2019}.  Since the \texttt{gdem} model is physically more realistic than a discrete 1 \texttt{cie} or 2 \texttt{cie} distribution \citep{vijayan2022} whereas the \texttt{wdem} model is better suited for cool-core clusters \citep[][]{werner2006,deplaa2006,mernier2016a}, we adopt it as our most plausible scenario.

\subsection{Instrumental cross-detection}

\begin{figure}
\centering
	\includegraphics[angle=0,width=0.47\textwidth, trim={0cm 0cm 0cm 0.8cm},clip]{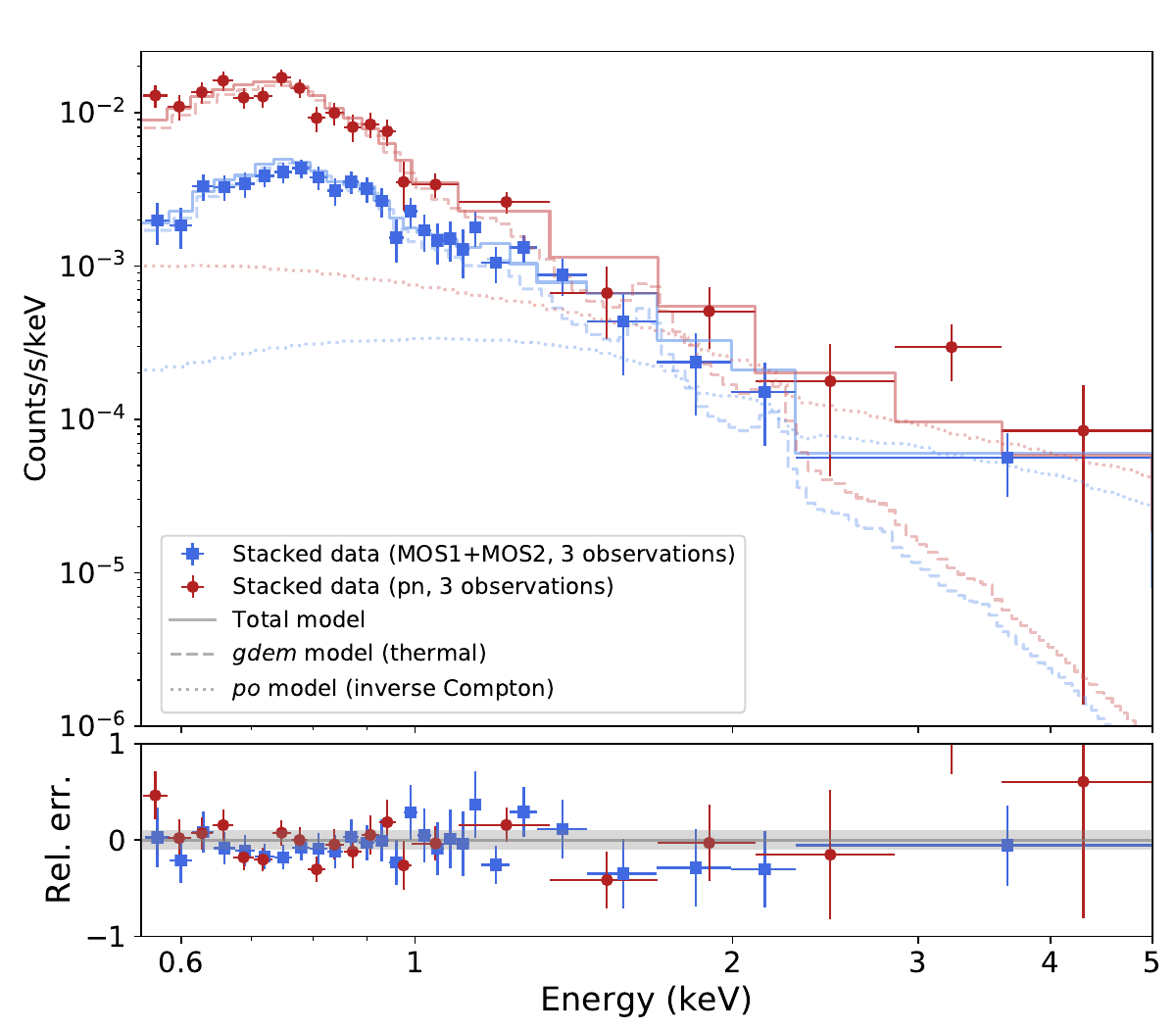}
\caption{Combined, background subtracted EPIC spectra of MRC\,0116+111 (\texttt{gdem} model), shown for MOS (1+2) and pn separately. For each instrument, the three pointings have been stacked and spectrally rebinned for clarity.
}
\label{fig:spectrum_MOSpn}
\end{figure}

Second, we must ensure that the residual power law component is not an instrumental artefact. To do so, we re-evaluate our free parameters for the MOS and pn spectra separately. Although we report a slight but significant tension in the gas temperature estimates from the two instruments ($kT_\mathrm{MOS} = (0.69 \pm 0.04)$~keV and $kT_\mathrm{pn} = (0.58 \pm 0.06)$~keV; likely due to cross-calibration imperfections of the Fe-L complex in the soft energy band), we find that the hard X-ray excess is detected with high significance using both instruments ($4.7\,\sigma$ and $5.4\,\sigma$ for MOS and pn, respectively). The detection of this excess is further demonstrated in Fig.~\ref{fig:spectrum_MOSpn}.

We have also carefully verified that our best-fitting results are left unchanged when we apply the recent corrections on the EPIC effective areas (available in SAS v20.0 and later) used for improving the cross-calibration between MOS and pn (XMM-CCF-REL-382)\footnote{https://xmmweb.esac.esa.int/docs/documents/CAL-SRN-0382-1-1.pdf} and between \xmm/EPIC and \nustar\ (XMM-CCF-REL-388)\footnote{https://xmmweb.esac.esa.int/docs/documents/CAL-SRN-0388-1-4.pdf}. Finding consistent results before and after applying this correction is not surprising since, at $E < 5$~keV (i.e. in our analyzed energy band), the difference between the effective areas is limited to $\lesssim$10\% and $\lesssim$2\%, respectively.

\subsection{Background uncertainties}\label{subsec:background}

Third, the detection of a hard spectral tail in our net EPIC spectra is in fact a differential test, as it means that the tail is present in the source region but not in the background region (also located on the same chip). In addition to the diffuse soft and hard X-ray emission components being spatially decoupled (Sect.~\ref{subsec:spatial}), this strongly suggests that background effects are unlikely to explain the hard excess. Nevertheless, photons falling beyond the cutoff of the thermal component (i.e. $E \gtrsim 2$~keV) are inevitably more likely to be confused with background counts (see also Fig.~\ref{fig:spectrum}) and, therefore, it is essential to ensure that the background is correctly accounted for in our spectral analysis. A first, encouraging indication is, as expected, the apparent absence of source emission in the spectra of the extracted background region. In fact, the \xmm\, background has been studied in-depth \citep[e.g.][]{kuntz2008,gastaldello2022} and, except a few fluorescent lines \citep[e.g.][]{mernier2015}, the shape of the hard particle background (HPB; dominating the hard energy band) is known to be relatively flat. In other words, a rescaling of that component should not alter the detectability of the (steeper slope) hard X-ray excess seen here. Nevertheless, the residual soft proton background (SPB) has a less predictable slope and a conservative modelling approach of the background with minimal assumption on the above two components is required as an extra safety check. Better suited for much more extended sources, details on this background-modelling approach are described in previous studies \citep{bulbul2012,mernier2015}. In summary, three astrophysical components are modelled together in addition to the instrumental HPB and SPB. The thermal foreground, originating from the local hot bubble and from the Galactic thermal emission, is then modelled with an (unabsorbed) \texttt{cie} model plus an (absorbed) \texttt{cie} model, with free emission measures but temperatures fixed to $kT_\mathrm{LHB} = 0.08$~keV and $kT_\mathrm{GTE} = 0.2$~keV, respectively. The cosmic X-ray background is modelled with a power law of index $\Gamma_\mathrm{CXB} = 1.41$ and an normalization of $3.7 \times 10^{49}$~photons~s$^{-1}$~keV$^{-1}$ as already estimated in Sect.~\ref{subsec:point_sources}. The two instrumental components, on the other hand, should be modelled without being folded through the ARF.  While the residual SPB is modelled with a simple power law, the HPB is modelled with a broken power law and a series of Gaussian components to account for the emission of fluorescent lines \citep{mernier2015}. The energy of these lines, as well as the slopes of the HPB continuum, are fixed by filter wheel closed observations \citep{mernier2015}. The free parameters of the two instrumental background components are therefore their normalizations (HPB and SPB) as well as the SPB slopes of MOS and pn spectra, all fitted separately in each of the three observations. This very conservative approach---with no a priori assumption from the rest of the field of view other than the CXB emissivity, inevitably results in many more degrees of freedom than in our initial methodology; nevertheless the power law component remains significantly detected ($3.6\,\sigma$) with similar best-fitting results. While this is reassuring, we emphasize again that in our case the use of local background is the most robust approach to detect subtle features such as the hard excess reported here.

\subsection{Energy range of the fit}

Fourth, whilst our adopted fitting range arguably constitutes the best compromise for maximizing the actual signal from the source while avoiding additional sources of noise, it is important to ensure that our results are not much affected by this choice, in particular by the upper energy limit as the main tracer of the non-thermal emission. Specifically, if the hard X-ray excess becomes background-dominated beyond certain energies, one expects the significance of our power law component to vary strongly as a function of the upper energy limit adopted in our fits, To check this, we refit our EPIC spectra setting the energy ranges to, successively, 0.55--4~keV and 0.55--10~keV. Encouragingly, this component remains confidently detected at similar flux densities regardless of the energy limit chosen for the spectral analysis.

\subsection{Atomic code uncertainties}

Fifth, the atomic code and data bases used to model the thermal emission of the gas (SPEXACT v3.06) originate from the state of the art calculations currently available. Other equally good codes exist, however, and could potentially yield a different spectral shape of the thermal component, eventually affecting the significance level of our hard X-ray detection. Using the XSPEC package to fit our thermal component with an \texttt{apec} model (using the up-to-date atomic database AtomDB v3.0.9), we find no noticeable change in our results. This is not surprising as previous work, at moderate spectral resolution, showed only minor differences in key parameters derived from these two models \citep{mernier2020,gastaldello2021}. For completeness, we repeat the same exercise using an outdated version of the SPEX atomic codes (SPEXACT v2.05), again with no impact on our conclusions.

\subsection{Uncertainties on absorption column density and metallicity}

Sixth, although thawing more parameters in our fits would inevitably lead to larger statistical uncertainties, it may be instructive to explore how doing so would affect our results. Encouragingly, we find that free $n_H$ and $Z$ still result in a fairly comfortable detection of the hard X-ray excess ($3.5\,\sigma$). In this case, we note that the best-fitting metallicity is found to drop to $\sim$0.07~Solar, which seems physically unrealistic given the $\gtrsim$30\%~Solar levels found in virtually all clusters and groups \citep{werner2013,mernier2018c}. In fact, fixing the metallicity to the most plausible values (i.e. between 0.5 and 1~Solar) results in stronger line emission, hence a steeper thermal profile and, in turn, further enhancing the significance of the non-thermal component.

\subsection{Slope of the non-thermal component}\label{subsec:gamfr}

Finally, beyond the strong reliability of the radio spectral index \citep{bagchi2009}, we also check whether the slope of the hard X-ray tail is indeed consistent with the value expected from its radio synchrotron emission (i.e. $\Gamma = 1.55$) as predicted in the IC emission scenario. This can be readily done by thawing the IC X-ray photon index $\Gamma$ in our simple \texttt{cie+po} test case or, even better, in our most realistic \texttt{gdem+po} case (Sect.~\ref{subsec:spectral}). Although the detection of the hard X-ray excess is left essentially unaltered (Table \ref{tab:systematics}), the slope of the \texttt{po} component significantly deviates from its original assumption, with $\Gamma$ values found to be ($2.87 \pm 0.25$) and ($2.6 \pm 0.4$) for the single- and multitemperature cases respectively. At first glance, this tension with $\Gamma = 1.55$ seems to disfavour the IC scenario as expected for this source. The reason, however could instead be found in the weights of the fitting statistics: at the expense of biasing its slope in the hard band, the fit uses the non-thermal component parameters to adjust small but significant residuals in the Fe-L complex of the thermal component in the soft band. This is demonstrated further in Appendix~\ref{app:gamfr} by estimating the best-fitting $\Gamma$ values while varying the energy range for the spectral fitting. In particular, we note that within 0.8--5~keV band, i.e. where the non-thermal model dominates and can be estimated with less bias, we find $\Gamma = (1.7_{-0.6}^{+0.7})$ and $\Gamma = (2.2 \pm 0.6)$ using respectively the \texttt{cie+po} and \texttt{gdem+po} modelling, thus in agreement with the slope expected from the IC scenario. Encouragingly, even in these extreme cases with much less available statistics, the non-thermal emission remains detected with $2.0\,\sigma$ and $2.4\,\sigma$ confidence for the \texttt{cie+po} and \texttt{gdem+po} modelling, respectively.

\section{Discussion}\label{sec:discussion}

\subsection{The spatial distribution of the hard X-ray component}\label{subsec:spatial_discussion}

In Sect.~\ref{subsec:spatial}, we have shown that the soft X-ray distribution---tracing the gas emission, and the hard X-ray distribution---tracing the hard excess seen in the spectra (Fig.~\ref{fig:spectrum}) are clearly distinguishable from each other spatially. Remarkably, the hard component exhibits two elongated peaks, the brighter one coinciding with the NE radio lobe remnant as well as with a possible cavity seen in the soft X-ray emission (Figs.~\ref{fig:EPIC} and \ref{fig:composite}). Although the second hard X-ray peak shows a $\sim$45~kpc offset from the centre of the SW radio lobe remnant, we note that the two hard peaks match in shape and extent. 
We also note that, at 2~keV, the combined non-thermal flux density of these two peaks is found to be $6.4 \pm 0.9 \times 10^{-14}$~erg~s$^{-1}$~cm$^{-2}$~keV$^{-1}$. This amounts to $(41 \pm 8)\%$ of the total non-thermal flux arising from our initial elliptical region ($1.55 \pm 0.22 \times 10^{-15}$~erg~s$^{-1}$~cm$^{-2}$~keV$^{-1}$). Investigating this further spectrally (Table~\ref{tab:systematics}), we circumscribe the double-peaked region with a polygon region (Appendix~\ref{app:regions}) and then find that the hard excess remains detected in both this region ($7\,\sigma$) and in the surrounding diffuse region ($5\,\sigma$; defined here as the entire group region minus the central double-peaked region). This result is important as it demonstrates that the detected hard X-ray tail is \textit{not} limited to just the above-mentioned two peaks: instead at least half of it is clearly diffuse and extends well beyond the two concentrated patches. Some possible implications of this diffuse IC scenario are developed in the next Section.

Overall, our tests described in the previous Section confirm that the hard X-ray tail detection reported in this work is robust to systematic uncertainties and unlikely to be of thermal origin. Broadly in line with the observed spatial decoupling of the soft X-ray emission from the diffuse hard X-ray component discussed in this section, the most natural interpretation by far would be in terms of a non-thermal, diffuse IC component originating from the galaxy group. In fact, simple models in which lobe remnants are filled with magnetic field and relativistic electrons naturally predict a spatial correlation between hard X-ray and radio components (also possibly coinciding with cavities sometimes observed within the thermal gas at softer energies) as indeed observed in the NE lobe remnant of MRC\,0116+111. Unexpected, however, is the spatial offset of the southern hard X-ray component from the SW lobe remnant. At first glance, it appears conceivable that an unusual plasma weather and dynamics within this group (for instance through vigorous sloshing motions) might have led to a localized decoupling of relativistic electron overdensities from regions of high magnetic field. In fact, we note that a slight anticlockwise rotation ($\sim$ 15--60$^{\circ}$) between the hard X-ray and radio components would result in a much better spatial correspondence between the two emission components, for the SW lobe remnant as well. Alternatively, the southern hard X-ray peak might represent a separate electron population. However, the spatial offset between these radio and X-ray components remains challenging to explain. Clearly, tailored magneto-hydrodynamical simulations as well as high-resolution radio observations at multiple frequencies are needed to further understand the behavior of relativistic electrons (and of their radio- and X-ray-related observables) in this group.

\subsection{Diffuse IC emission and magnetic field estimates}\label{subsec:implications}

There are a number of claimed detections of IC X-ray emission from ``hot spots'' seen in radio-galaxies \citep[e.g.][]{harris1994,hardcastle2002} and from collimated jets of relativistic particles in blazars \citep[e.g.][]{worrall2020,hodges2021}, although such an interpretation for the latter objects has been questioned in some cases \citep[e.g.][]{jester2006,uchiyama2007,cara2013,breiding2017}. Regardless of this, the present finding differs both in the nature of the system and the physical scale and kinematics of the IC emitting plasma. Whilst our hard X-ray image reveals a non-thermal contribution from the vicinity of the likely radio lobe remnants associated with the central dominant galaxy (see Fig.~\ref{fig:EPIC}), the flux contributed by these peaks accounts for less than half of the total IC emission detected from the entire volume of the group, such that the surrounding diffuse region exhibits a non-thermal X-ray excess too (Sect.~\ref{subsec:spatial_discussion}). To our knowledge, this is the first time that \textit{extended} IC emission, associated both with relativistic electrons pervading the entire group and also from a more confined, double-peaked region, has been simultaneously detected in a system containing multiple galaxies.

We propose that both the diffuse radio emission and the diffuse IC X-ray emission share a common origin from the same population of relativistic electrons permeating large scale structures. Whereas the emissivity of the former depends on \textit{both} the relativistic electron density \textit{and} the volume-averaged magnetic field (hence implying a degeneracy between these two \textit{a priori} unknown values), the emissivity of the latter depends just on a single unknown, namely the relativistic electron density. Consequently, the ratio of the radio and IC X-ray flux densities provides unique constraints on the magnetic field strength \citep[e.g.][]{feretti2012,ota2014,mernier2019}, free from the biases that plague other methods (Sect.~\ref{sec:intro}). In fact, upper limits to IC X-ray emission from previous work on clusters have so far provided only lower limits on intracluster magnetic fields of $\gtrsim 0.1-1~\mu$G \citep{bartels2015,cova2019,rojas2021}. 

Our present detection of IC emission allows accurate and essentially model independent estimate of the volume-averaged magnetic field pervading the group. For the entire group volume, we find the direct estimate to be $(1.9 \pm 0.3)~\mu$G. Considering separately the central double-peaked region and the surrounding diffuse hard X-ray region---and assuming no substantial variation of the radio spectral shape across the group \citep{bagchi2009}, we find average magnetic field estimates of $(1.38 \pm 0.14)$ and $(1.7 \pm 0.3)~\mu$G within and outside the hard X-ray peaks, respectively, which are statistically indistinguishable. While cosmological magneto-hydrodynamical simulations do predict $\mu$G level magnetic fields in the cores of rich clusters, they suggest fields that are smaller by one or two orders of magnitude in groups with gas densities and temperatures similar to MRC 0116+111 \citep[e.g.][]{dolag2002,donnert2018}. Interestingly, were the magnetic field in our object indeed so weak, the corresponding diffuse IC X-ray emission would be a factor $\gtrsim 30$ stronger than even our detection reported here. Whereas \nustar\ observations could in principle enable even more robust characterization of IC X-ray emission of this source, targeted searches using current and future observing facilities will be vital for ascertaining whether the present object has a magnetic field that is representative of a wider population of groups. This will be attempted by taking advantage of (i) synergies between X-ray surveys (such as eROSITA) and radio surveys (such as LOFAR, JVLA or SKA; see e.g. Richard-Laferri\`{e}re et al., in preparation), and (ii) the outstanding spectral and/or photon collecting capabilities offered by the next generation of X-ray observatories (e.g. \xrism, \athena).

\section{Conclusions}\label{sec:conclusion}

In this work, we have made use of very deep \xmm\ EPIC observations of the galaxy group MRC\,0116+111, hosting bright diffuse radio synchrotron emission. This source displays an exceptionally high radio-to-X-ray flux ratio along with a relatively low gas temperature ($kT \lesssim 0.7$~keV). The combination of these two attributes makes the source a prime target for undertaking a search for diffuse IC X-ray emission, not yet detected robustly in large scale systems.

Our analysis reveals a significant excess of hard X-ray emission (i.e. above $\sim$2~keV), which persists even accounting for the uncertainties arising from multitemperature thermal plasma and uncertainties in background subtraction, instrumental cross-calibration, plasma codes and spectral modelling. Explaining such spectral shape with pure thermal modelling would imply astrophysical conditions never witnessed in galaxy clusters/groups so far. Instead, the observed spatial distribution of this diffuse hard X-ray component---which is spatially distinct from the soft, thermal emission but shows a close spatial correspondence with the radio synchrotron emission---strongly favouring IC as the most likely source of the observed hard X-ray emission. This close but imperfect spatial correspondence of the hard X-ray peaks and radio lobe remnants, however, remains to be fully understood.

Following this interpretation, for the first time, we were able to constrain the volume-averaged magnetic field of a galaxy group in an unbiased manner. Our estimate of $(1.9 \pm 0.3)~\mu$G is in line with magnetic field intensities expected in the cluster regime, however it is 1--2 orders of magnitude larger than predicted for groups. Future radio and X-ray observations will be essential to better characterize the nature of this source and refine our present results, while specific magneto-hydrodynamical simulations are needed to better understand the dynamical and magnetic processes at play in this rather extreme system.

\section*{Acknowledgements}

We thank the anonymous referee for constructive comments that helped to improve this manuscript. FM thanks Soumyajit Mandal for fruitful discussions. The material is based upon work supported by NASA under award number 80GSFC21M0002. NW is supported by the GACR grant 21-13491X. M-LG-M acknowledges financial support from grant RTI2018-096228-B-C31 (MCIU/AEI/FEDER,UE), from the coordination of the participation in SKA-SPAIN financed by the Ministry of Science and Innovation (MICIN), and from the State Agency for Research of the Spanish Ministry of Science, Innovation and Universities through the `Center of Excellence Severo Ochoa' awarded to the Instituto de Astrof\`{i}sica de Andaluc\`{i}a (SEV-2017-0709). G-K thanks the Indian National Science Academy for a Senior Scientist position. A.R.L. is supported by the Gates Cambridge Scholarship, by the St John's College Benefactors' Scholarships, by NSERC through the Postgraduate Scholarship-Doctoral Program (PGS D) under grant PGSD3-535124-2019 and by FRQNT through the FRQNT Graduate Studies Research Scholarship - Doctoral level under grant \#274532. AS is supported by the Women In Science Excel (WISE) programme of the Netherlands Organisation for Scientific Research (NWO), and acknowledges the World Premier Research Center Initiative (WPI) and the Kavli IPMU for the continued hospitality. This work is based on observations obtained with \textit{XMM-Newton}, an ESA science mission with instruments and contributions directly funded by ESA member states and the USA (NASA). SRON is supported financially by NWO.

\section*{Data Availability}

The original data presented in this article are publicly available from the \textit{XMM-Newton} Science Archive (https://nxsa.esac.esa.int/nxsa-web/). The corresponding \textit{XMM-Newton} data are reduced with the SAS software (https://www.cosmos.esa.int/web/xmm-newton/sas) and further analyzed spectrally with the SPEX fitting package (https://www.sron.nl/astrophysics-spex). Additional derived data products can be obtained from the main author upon request.



\bibliographystyle{mnras}
\bibliography{MRC0116} 




\appendix

\section{Relevant spatial regions}\label{app:regions}

\begin{figure}
\centering
	\includegraphics[angle=0,width=0.49\textwidth, trim={0cm 0cm 0cm 0cm},clip]{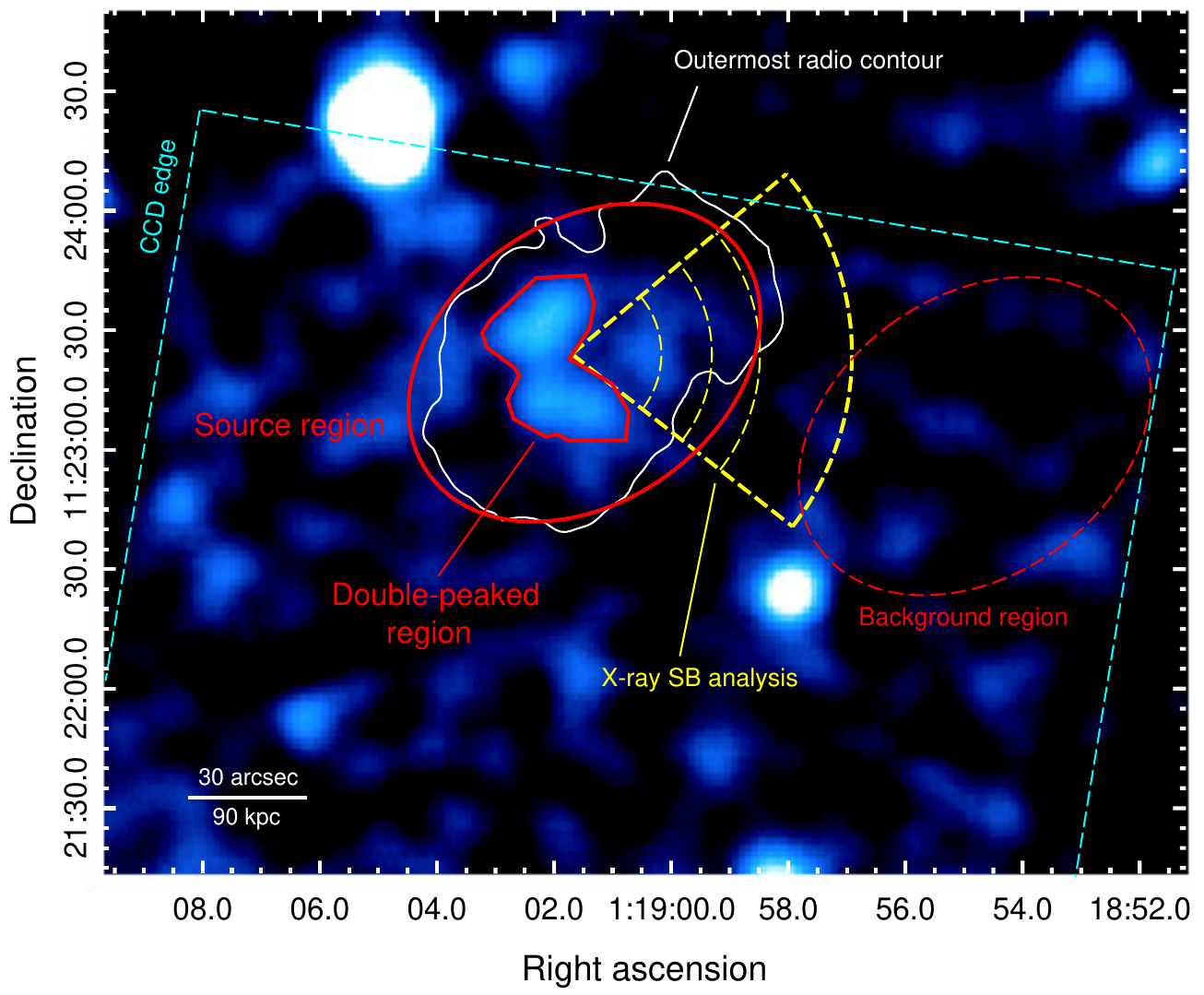}
\caption{Combined EPIC image of MRC\,0116+111 in the 2--4.5~keV band (smoothed and uncorrected from exposure maps for better visibility of the CCD edges near the source), annotated with all regions relevant to this work. For clarity, only the the outermost radio contour is displayed. 
}
\label{fig:regions}
\end{figure}

For the sake of clarity, we show in Fig.~\ref{fig:regions} a combined, smoothed EPIC image in the hard energy band (2--4.5~keV) annotated with all regions that are relevant to this work. These regions are: (i) the source and the background regions, both used in our spectral analysis over the whole group extent (Sect.~\ref{subsec:spectral}); (ii) the `central' double-peaked region identified in Sect.~\ref{subsec:spatial} and analyzed further in Sect.~\ref{subsec:spatial_discussion} (the surrounding `diffuse' region being simply defined as the source region minus the double-peaked region); and (iii) the wedge region used for our soft and hard SB analysis of the diffuse emission (see Sect.~\ref{subsec:spatial}). We also show the outermost contour of the radio emission (621 MHz, GMRT), omitting the inner contours for clarity. All these regions belong to the same CCD chip of our EPIC observations.

\section{Fitting systematics on the estimate of the non-thermal spectral slope}\label{app:gamfr}

\begin{figure*}
\centering
	\includegraphics[angle=0,height=0.29\textheight, trim={0.3cm 0.2cm 0cm 0.4cm},clip]{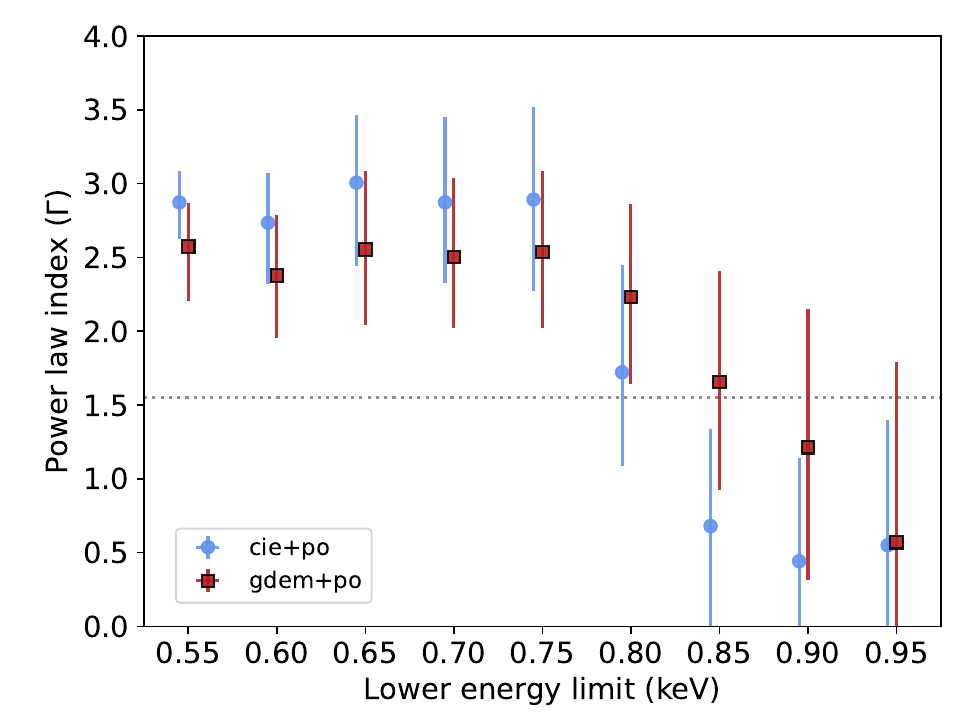}
	\includegraphics[angle=0,height=0.29\textheight, trim={0cm 0cm 0cm 0.5cm},clip]{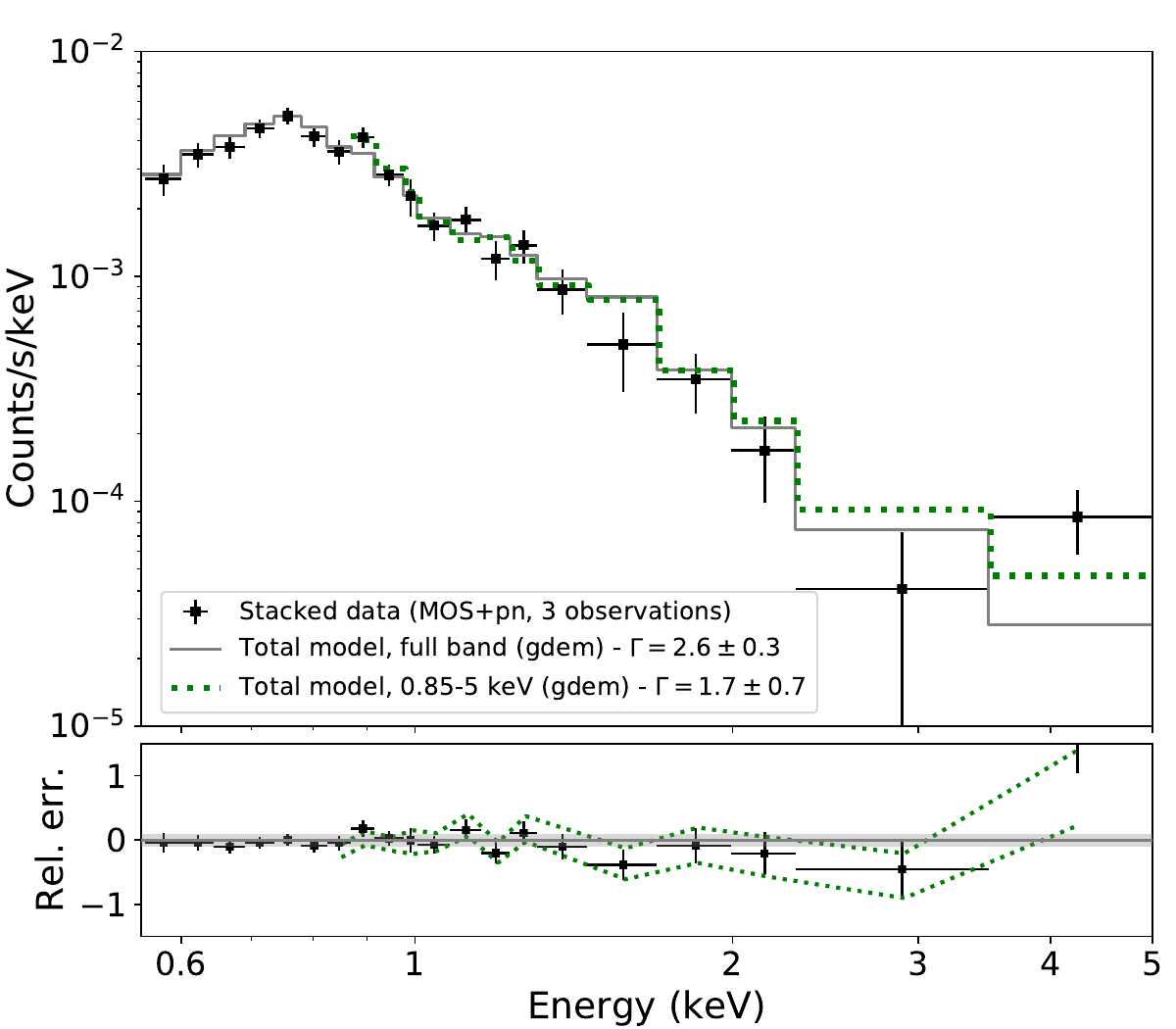}
\caption{\textit{Left:} Best-fitting estimates of the slope $\Gamma$ of the \texttt{po} model and its dependency on the lower energy limit over the considered fitting. The (mean) temperature of the thermal (i.e. \texttt{cie} or \texttt{gdem}) component is fixed to its best-fitting value obtained over the full fitting range (0.55--5~keV) with free $\Gamma$ (i.e. the leftmost data points on the figure).  The $\Gamma$ index expected from IC emission (i.e. adopted from the radio spectra index at low frequencies) is shown by the dashed grey line. \textit{Right:} Combined EPIC spectrum fitted with $\Gamma$ as free parameter, successively within the full band (grey histograms and black data points in the upper and lower panel, respectively) and within 0.85--5~keV (green dotted lines in both panels).
} 
\label{fig:gamfr}
\end{figure*}

As reported in Sect.~\ref{subsec:gamfr}, freeing the photon index parameter $\Gamma$ of the power law component in our initial broad-band \texttt{gdem+po} (or \texttt{cie+po}) modelling leads to an estimate that is in tension with the value expected from the IC scenario (i.e. $\Gamma = 1.55$). Such a tension, however, can be explained as an artefact arising from a biasing of the $\Gamma$ estimate towards counts in the soft band (much less relevant here, as it is dominated by thermal photons). To demonstrate this effect, we repeat the same exercise as in Sect.~\ref{subsec:gamfr} now varying successively the lower energy limit ($E_\text{low}$) of the fitting range. Since the (mean) temperature of the thermal component cannot be constrained when $E_\text{low} \gtrsim 0.7$~keV, we keep it fixed to its best-fitting value obtained in the full band fitting (i.e. for $E_\text{low} = 0.55$~keV)\footnote{For consistency, however, we have further verified that the best-fitting (mean) temperature is mostly unchanged as long as the Fe-L complex remains at least partly included in the fit (i.e. for $0.55~\text{keV} \le E_\text{low} < 0.7~\text{keV}$).}. As seen from Fig.~\ref{fig:gamfr} (left panel), the best-fitting $\Gamma$ estimate progressively approaches the expected value of 1.55 when $E_\text{low} \gtrsim 0.75$~keV, i.e. as soon as the power law component begins to dominate the considered fitting range. Such a difference can be easily explained as the fit systematically tends to reproduce the bright---hence `precise'---Fe-L complex with much higher priority than the fainter hard X-ray excess. Precision, however, often differs from accuracy: it is, in fact, well known that current thermal models (and their sometimes complicated temperature structure) cannot yet perfectly reproduce the complicated shape of the Fe-L complex \citep[e.g.][]{mernier2022,gu2022}. Within a broad-band, the fit will thus use the freedom to vary $\Gamma$ as a way to (incorrectly) adjust to the shape of the Fe-L complex. As opposed to this, setting a higher $E_\text{low}$ limit allows a less biased estimate of $\Gamma$, as the fitting process is now weighted more towards the power law component itself. In fact, the lack of substantial difference between the broad-band and hard-band best-fitting models seen in Fig.~\ref{fig:gamfr} despite their different $\Gamma$ best-fitting estimates (right panel) further argues that, despite its small \textit{statistical} uncertainties over the broad-band fit, the non-thermal spectral slope remains genuinely difficult to constrain and is, by definition, consistent with its theoretical 1.55 value. 

\bsp	
\label{lastpage}
\end{document}